\documentclass[11pt]{article}
\usepackage{geometry}                % See geometry.pdf to learn the layout options. There are lots.
\usepackage{graphicx}
\usepackage{hyperref}
\usepackage{amsmath, amssymb}
\usepackage{epstopdf}
\usepackage{color}

\usepackage{bm}%bold faced

\newcommand{\supp}{\mathrm{supp}}
\title{Quantum field theory on curved spacetime and the standard cosmological model}
\author{Klaus Fredenhagen and Thomas-Paul Hack}
\begin{document}

\maketitle

\begin{abstract}The aim of this review is to outline a full route from the fundamental principles of algebraic quantum field theory on curved spacetime in its present-day form to explicit phenomenological applications which allow for comparison with experimental data. We give a brief account on the quantization of the free scalar field and its Wick powers in terms of an algebra of functionals on configuration space. Afterwards we demonstrate that there exist states on this algebra in which the energy momentum tensor is qualitatively and quantitatively of the perfect fluid form assumed in the standard model of cosmology up to small corrections. We indicate the potential relevance of one of these corrections for the actively debated  phenomenon of Dark Radiation.
\end{abstract}

\section{Introduction}
The attempt to incorporate gravity into quantum theory meets great conceptual difficulties. The main reason for these problems seems to be the rather different roles played by space and time in quantum theory and in Einstein's theory of gravity. In quantum theory, an a priori notion of space and time enters the formulation and the interpretation of the theory in a crucial way. In Einstein's theory of gravity, on the other side, the structure of space and time is dynamical and strongly influenced by  the distribution of matter which is treated classically. 

These severe conceptual problems are accompanied by hard technical problems, hence testing ideas for solving the problem turns out to be extremely time consuming, and it is difficult to obtain reliable conclusions. In despair, rather radical approaches have been proposed as e.g. string theory and loop quantum gravity, but we think that it is fair to say that none of these approaches has reached its goal, up to now, nor could either of them be ruled out, neither by empirical results nor by inner theoretical reasons.

If one is less ambitious and takes into account, that gravitational forces tend to be very small compared to other forces, one   
may consider, in a first step, gravity as an external field, producing a curved spacetime, and treat quantum matter by quantum field theory on such a background.  
One may then, in a second step, treat quantum gravity as a quantum field fluctuating around a given background.

The second step meets severe problems: the arising theory is nonrenormalizable, which means that in every order of perturbation theory new interaction terms appear whose coupling constants have to be determined by experiments. Moreover, the causal structure of the theory is determined by the background metric, whereas physics would require that it depends only on the full metric, including the quantum fluctuations. Nevertheless, a consistent perturbative formulation
was recently presented by Brunetti, Rejzner and one of us in \cite{Brunetti:2013maa}. 

Surprisingly, already the first step is by no means trivial. The reason is, that quantum field theory in its standard formulation heavily depends on the symmetries of Minkowski space. These symmetries are used to define the vacuum and the concept of a particle, and one can then, under quite general conditions, derive the existence of scattering states and of an S-matrix. 

But on a generic Lorentzian spacetime, no nontrivial symmetries exist, and as a consequence, neither the concept of a vacuum state nor that of particles can be   
intrinsically introduced. In particular, the classical picture of particles moving in an empty  spacetime is not supported by quantum field theory. The most spectacular consequence of this fact is the evaporation of black holes as predicted by Hawking.   

The problems of quantum field theory on a given curved back ground have been solved within the last 20 years by using the
concepts of algebraic quantum field theory and by replacing techniques of operators on Fock space by methods from microlocal analysis \cite{Hormander}. A compilation of references on algebraic quantum field theory on curved spacetimes can be found in \cite{Benini:2013fia}.

Algebraic quantum field theory was originally developed in order to understand the relation between the local degrees of freedom of quantized fields and the observed multi-particle states \cite{Haag:1992hx}. It was then observed by Dimock and Kay that it provides a good starting point for formulating a theory on a curved spacetime \cite{Kay:1979xi, Dimock}. 
The absence of a distinguished Hilbert space representation, however, was a severe obstacle for extending the theory to nonlinear fields, 
the most prominent being the energy momentum tensor.  

For this purpose it was necessary to understand the singularities of correlation functions. There was overwhelming evidence that the so-called Hadamard states yield a class of states with the correct singularity structure. A direct characterization of Hadamard states turned out to be rather complicated \cite{Kay:1988mu}, and its use for the determination of correlation functions of nonlinear fields seemed to be extremely cumbersome. 

The situation changed completely when Radzikowski discovered that the Hadamard condition could equivalently be replaced by a positivity condition on the wave front set of the 2-point function \cite{Radzikowski:1996pa, Radzikowski:1996ei}. This marked the breakthrough for the modern theory of quantum fields on curved back grounds, and within a few years it was possible to construct all kinds of composite fields \cite{Brunetti:1995rf} and to prove the existence of renormalized time ordered products \cite{Brunetti:1999jn}.

Renormalization, however, had still the problem that renormalization conditions at different points of spacetime could not be compared with each other in the absence of nontrivial symmetries. A new principle was needed, the principle of local covariance \cite{Brunetti:2001dx}. This principle says that it is not meaningful to do physics on a special spacetime; instead all structures should depend only on the local geometry. Based on this principle, Hollands and Wald were able to finish the renormalization program \cite{Hollands:2001nf, Hollands:2001fb}, which had been started by Brunetti and one of us \cite{Brunetti:1999jn}. One of the outcomes of this generalization of algebraic quantum field theory is that it is meaningful to consider the same field on different spacetimes.

A direct application of this fact is the use of the energy momentum tensor as a source term for Einstein's equation. But as long as gravity itself is not quantized
one has the problem to compare a quantum object with a classical object. On a pragmatic level this may be solved by using the expectation value of the energy momentum tensor. This might be reasonable as long as the fluctuations are small enough. But here new problems arise. One is the fact that the correlation functions of the energy momentum tensor diverge at coinciding points. One therefore looks at appropriate averages; this, however, introduces a new parameter into the theory. The other problem is even worse: whereas fields exist which can be considered to be the same on different spacetimes, a corresponding identification of states on different spacetimes does not exist.        

The latter problem can presumably only be treated in a theory containing quantized gravitational and matter fields. One may, however, restrict oneself to situations 
with higher symmetries, as they arise in cosmological spacetimes of the Friedmann-Robertson-Walker type. There, one may admit only states which are invariant
under the spatial symmetries. Still, this does not fix the states uniquely, hence additional choices have to be introduced. Nevertheless, one can in this way reproduce the standard cosmological model from first principles, by modelling the matter-energy content of the universe entirely in terms of quantum fields rather than effectively by means of a classical perfect fluid \cite{Hack:2013aya}. 

\section{The free scalar field and its normal ordered products}
Classically, a configuration of a scalar field may be understood as a smooth function on spacetime. Let $\mathcal{C}^{\infty}(M)$ be the set of all smooth functions on a spacetime $M$, and let $\mathrm{Sol}(M)$ be the subset of smooth solutions of the Klein-Gordon equation.
Classical observables are functions on $\mathcal{C}^{\infty}(M)$ modulo functions which vanish on solutions. The observables of the quantum theory form a suitable subspace on which the algebraic structures of quantum theory can be defined. This subspace can be characterized in the following way. 

We consider a globally hyperbolic time oriented spacetime. On such a spacetime the Klein-Gordon equation
$$P\phi=\left(\nabla^a \nabla_a + \xi R + m^2\right) \phi=0\,,$$
with curvature scalar $R$, curvature coupling parameter $\xi$ and mass $m$, 
 possesses unique retarded and advanced Green's functions $\Delta_{R,A}$ considered as maps from compactly supported densities to smooth functions. Their difference is the commutator function $\Delta$. A Hadamard solution of the Klein-Gordon operator $P$
is a distributional bisolution $h$ with
the properties:
\begin{enumerate}
\item $h(x,y)-h(y,x)=i(\Delta(x,y))$
\item $\mathrm{WF}(h)=\{(x,x';k,k')\in \mathrm{WF}(\Delta)|k\in V^+_x\}$ where $V^+_x$ is the closed forward lightcone in $T^*_xM$.
\item $h$ is a distribution of positive type.
\end{enumerate} 
We want to introduce an associative product $\star_h $ on a subspace $\mathcal{F}(M)$ of the space of maps $\{F:\mathcal{C}^{\infty}(M)\to\mathbb{C}\}$ by setting
\begin{equation}
(F\star_h G)(\phi)=\sum_{n=0}^{\infty}\frac{\hbar^n}{n!}\left\langle \frac{\delta^nF}{\delta\phi^n},h^{\otimes n}\frac{\delta^nG}{\delta\phi^n}\right\rangle(\phi)\ .
\end{equation} 
In order to make this definition meaningful we require for $F\in\mathcal{F}(M)$:
\begin{enumerate}
\item $F$ is polynomial, therefore the sum over $n$ is finite.
\item $F$ is smooth in the sense of the calculus on locally convex spaces, where $\mathcal{C}^{\infty}(M)$ is equipped with its standard topology (uniform convergence of all derivatives on any compact set).
From these two conditions it follows that $F$ is of the form
\[F(\phi)=\sum_{n=0}^{N}\langle f_n,\phi^{\otimes n}\rangle \]
with compactly supported distributional densities $f_n$ on $M^n$.
\item The wave front set of $f_n$ does not intersect $(V^+)^n$ nor $(V^-)^n$. This condition guarantees by H\"ormander's theorem on the multiplicability of distributions, that, in view of the wave front set of the Hadamard solution, the summands in the definition of the product are well defined. 
\end{enumerate}
The product is associative. Complex conjugation induces an involution on $\mathcal{F}(M)$,
\[\overline{F}\star_h\overline{G}=\overline{G\star_h F}\ ,\]
hence $\mathcal{F}(M)$ gets the structure of a unital *-algebra, where the unit is the constant function $F(\phi)\equiv1$. The subspace $\{F\in\mathcal{F}(M)|F(\phi)=0\text{ for }\phi\in\mathrm{Sol}(M)\}$ is an ideal, and the quotient is the enlarged CCR-algebra. It contains as a subalgebra the CCR-algebra generated by linear functionals of the form $F(\phi)=\langle f,\phi\rangle$ with a smooth density $f$ on $M$ and in addition all local polynomials in the field and its derivatives,
\[F(\phi)=\int f(j_x(\phi))d\mathrm{vol}(x)\]
where $x\mapsto j_x(\phi)=\{\phi+\psi|\psi\in\mathcal{C}^{\infty}(M)\text{ with }\partial^{\alpha}\psi(x)=0\text{ for all multiindices }\alpha\}$ is the jet prolongation of $\phi$, and 
$f$ is a smooth function on the jet bundle which is a polynomial in $\phi$ and its derivatives at every point $x\in M$, and which has compact spacetime support
\[\supp F=\overline{\bigcup_\phi\supp(x\mapsto f(j_x(\phi))}\ .\] 

The definition of the enlarged CCR-algebra depends on the choice of the Hadamard solution $h$. Since two Hadamard solutions differ by a smooth symmetric and real valued bisolution $w$, the arising algebras are isomorphic with the isomorphism
\[\Gamma_w=\exp{\frac12\hbar\left\langle w,\frac{\delta^2}{\delta\phi^2}\right\rangle}\ .\] 

Every Hadamard solution $h+w$ induces a family of coherent states by
\[\omega_{w,\phi}(F)=(\Gamma_wF)(\phi)\]
with $\phi\in\mathrm{Sol}(M)$. According to a result of Verch, the arising GNS-representations are locally equivalent \cite{Verch:1992eg}.

A further crucial ingredient for the interpretation of the theory are locally covariant fields $A$. These are, for every spacetime $M$,  linear maps $A_M$ from the space of (compactly supported) test tensors to the algebra $\mathcal{F}(M)$ such that, for every isometric, time orientation and causality preserving embedding $\chi:M\to N$
into a larger spacetime $N$ one has the relation
\[A_M(f)(\phi\circ\chi)=A_N(\chi_*f)(\phi)\]
where $\chi_*$ denotes the push forward of test tensors.  In other words, a locally covariant field is a natural transformation between the functor $\mathcal{D}$ of test tensor spaces and the functor $\mathcal{F}$ of observable algebras, both based on the category of globally hyperbolic spacetimes with isometric, time orientation and causality preserving embeddings as morphisms.

In a first attempt one may look at a polynomial $p(\partial^{\alpha}\phi,\alpha\in\mathbb{N}_0^d)$ in $\phi$ and its derivatives and set
\[A_M(f)(\phi)=\int f(x)p(\partial^{\alpha}\phi(x))d\mathrm{vol}(x)\ .\] 
But this definition violates the naturality condition for locally covariant fields since there is no natural choice for the Hadamard solution, i.e. no choice which is compatible with all possible embeddings of a spacetime into another one, a fact which is responsible for the nonexistence of a vacuum state.

Let $p(\nabla)$ be a polynomial in covariant derivatives (with respect to the Levi-Civita connection) and consider the functionals
\[A(x)(\phi)=e^{p(\nabla)\phi(x)}\ .\]
Under the isomorphism $\Gamma_w$, $A(x)$ transforms as
\[\Gamma_wA(x)=e^{\frac{1}{2} p(\nabla)\otimes p(\nabla)w(x.x)}A(x)\ .\]
We now use the fact, that $w$ is the difference of 2 Hadamard solutions. Hadamard solutions admit an asymptotic expansion
\begin{align*}h(x,y)&=\frac{u(x,y)}{\sigma(x,y)}+\sum_{n=0}^Nv_n(x,y)\sigma(x,y)^n\ln(\mu^2\sigma(x,y))+w^h_N(x,y)\\
&= h^\text{sing}_N(x,y)+w^h_N(x,y)\ .\end{align*}
Here $x,y$ are points in a geodesically convex open set, $\sigma(x,y)$ is the signed square of the geodesic distance between $x$ and $y$,  the functions $u$ and $v_n$ are solutions of the so-called transport equations and are uniquely determined by the local geometry. $\mu$ is a free parameter with the dimension of inverse length. $w^h_N$ is an $2N+1$ times continuously differentiable function which depends on the choice of $h$. We omit the $\epsilon$-prescription necessary for $h^\text{sing}_N$ to be well-defined, see \cite{Kay:1988mu}.

We now set
\[A_h(x)=e^{\frac12p(\nabla)\otimes p(\nabla)w_N^h(x,x)}A(x)\]
where $N$ is larger than or equal to twice the degree of $p$, and find
\[\Gamma_{h-h'}A_{h'}(x)=A_h(x)\ .\]
By expanding the exponential series we obtain a large class of locally covariant fields. These correspond to Wick powers of the scalar field and its derivatives regularised by point-splitting and suitable subtractions of derivatives of $h^\text{sing}_N$. This class may be enlarged by the $\phi$-independent locally covariant fields constructed from the metric. Further details may be found e.g. in \cite{Fredenhagen:2011mq}.
%\subsection{}

A locally covariant field of particular interest is the energy momentum tensor $T_{ab}(x)$. However, it is by no means intrinsically clear which locally covariant field is the observable whose expectation value is the ``correct'' source term for Einstein's equation. Essentially this is due to the fact that gravity is sensitive to absolute energy densities rather than energy density differences. Wald \cite{Wald:1995yp} and later Hollands and Wald \cite{Hollands:2004yh} have suggested that a locally covariant field should satisfy standard commutation relations, covariant conservation $\nabla^aT_{ab}(x)(\phi)=0$ and suitable analyticity conditions in order to be a meaningful energy momentum tensor. For a free scalar field this implies that the most general energy momentum tensor is of the form
\begin{equation}\label{eq_nonunique}
T_{ab}(x)(\phi) = T^0_{ab}(x)(\phi) + \alpha_1 g_{ab}(x) + \alpha_2 G_{ab}(x) + \alpha_3 I_{ab}(x) + \alpha_4 J_{ab}(x)\,,
\end{equation} 
where $G_{ab}$ is the Einstein curvature tensor whereas $I_{ab}$ and $J_{ab}$ are local curvature tensors which are obtained as functional derivatives with respect to the metric of the action functionals $\int\sqrt{-g}R^2 d\text{vol}(x)$ and $\int\sqrt{-g}R_{ab}R^{ab}d\text{vol}(x)$ respectively. Moreover, a possible ''model'' $T^0_{ab}$ is the functional
\begin{equation}\label{eq_deftmunu}T^0_{ab}(x)(\phi)=T^\text{class}_{ab}(x)(\phi)+\lim\limits_{x\to y} \left(D_{ab}-\frac13 g_{ab} P_x\right)w^h_N(x,y)\qquad N\ge 1
\end{equation}
where $T^\text{class}_{ab}$ is the classical energy momentum tensor of the scalar field, $D_{ab}$ is a second order bi-differential operator defined by $\lim_{x\to y}D_{ab}w(x,y)=\langle w,\frac{\delta^2}{\delta\phi^2}\rangle T^\text{class}_{ab}(x)(\phi)$ and the modification term $-\frac13 g_{ab} P_x$ is necessary in order to have a covariantly conserved $T^0_{ab}$ \cite{Mo03}. The four parameters $\alpha_i$ are free parameters which can not be determined intrinsically within QFT on curved spacetimes, but only by measurements or within a more fundamental theoretical framework.

An alternative ''model'' $T^0_{ab}$ can be obtained by taking the functional derivative with respect to the inverse metric of the ``one-loop effective Lagrangean''
$${\mathcal L}^0(\phi)(x)={\mathcal L}^\text{class}(\phi)(x)+\left\langle w^{h_{dWS}}_N,\frac{\delta^2}{\delta\phi^2}\right\rangle{\mathcal L}^\text{class}(\phi)(x)\qquad N\ge 1.$$
Here $w^{h_{dWS}}_N$ is the regular part of the deWitt-Schwinger Hadamard solution $h_{dWS}$ which is a formal series in $\sigma(x,y)$ with purely geometric coefficients \cite{Hack:2012dm2}.

\section{The standard cosmological model in quantum field theory on curved spacetimes}
\label{sec_cosmology} 

In the standard cosmological model the universe is modelled by a Friedmann-Lema\^itre-Robertson-Walker (FLRW) spacetime $(M,g)$ with manifold $M=I \times {\mathbb R}^3\subset {{\mathbb R}^4}$ and metric $g=dt\otimes dt - a^2(t) dx^i\otimes dx_i$. We consider the case where the spatial slices are diffeomorphic to ${\mathbb R}^3$ for simplicity and because this is favoured by observations. Here $t$ is cosmological time, whereas the scale factor $a(t)$ is a smooth non-negative function whose logarithmic $t$-derivative is the Hubble rate
$H$, which is assumed to be strictly positive in what follows. Further convenient time variables are the conformal time $\tau$, the scale factor $a$ and the redshift $z:=a_0/a-1$, where $a_0=1$ is the scale factor of today. These time variables are related by $dt = a d\tau = \frac{da}{a H}= -\frac{dz}{(1+z)H}\,$. 

Given the high symmetry of $(M,g)$ and the Einstein equation $G_{ab}=8\pi T_{ab}$, the energy momentum tensor $T_{ab}$ must be of perfect fluid form and thus determined by the energy density $\rho=(\partial_t)^a(\partial_t)^b T_{ab}$ and pressure $p$, which are related by the equation of state $p=p(\rho)$. Moreover, the Einstein equation is equivalent to the (first) Friedmann equation
$$H^2 = \frac{8\pi G}{3}\rho$$
and a conservation equation. According to the standard model of cosmology -- the $\Lambda$CDM-model -- our universe contains matter, radiation, and Dark Energy, modelled macroscopically as perfect fluids with equation of state $p = w \rho$, $w=0, \frac13, -1$ for matter, radiation and Dark Energy (assuming that the latter is just due to a cosmological constant) respectively. Consequently, the Friedmann equation can be conveniently rewritten as 
\begin{equation}\label{eq_friedmannlcdm}
\frac{H^2}{H^2_0}=\frac{\rho_\text{$\Lambda$CDM}}{\rho_0}=\Omega_\Lambda + \frac{\Omega_m}{a^3} + \frac{\Omega_r}{a^4}\,,\quad \rho_0 = \frac{3 H^2_0}{8\pi G}\,,
\end{equation}
where $H_0$ is the present Hubble rate -- the Hubble constant -- and the constants $\Omega_\Lambda$, $\Omega_m$, $\Omega_r$ denote the present fractions of the energy density due to Dark Energy, matter and radiation respectively. Observations indicate approximately \begin{equation}\label{eq_omegas}\Omega_m= 0.3,\quad \Omega_r = 10^{-4},\quad\Omega_\Lambda=1-\Omega_m-\Omega_r\end{equation}
see \cite{Ade:2013zuv} for the latest exact values from the Planck collaboration.
In the context of cosmology the terms "matter" and "radiation"  subsume all matter-energy with the respective macroscopic equation of state such that e.g. "radiation" does not encompass only electromagnetic radiation, but also the three left-handed neutrinos present in standard model of particle physics (SM) and possibly so-called Dark Radiation,  and "matter" subsumes both the baryonic matter which is in principle well-understood in the SM and Dark Matter. Here, Dark Matter and Dark Radiation both quantify contributions to the macroscopic matter and radiation energy densities which exceed the ones expected from the knowledge of the SM and are believed to originate either from fields not present in the SM or from other sources, i.e. modifications of classical General Relativity. 

Notwithstanding, at least the contributions to the macroscopic matter and radiation energy densities which are in principle well-understood originate microscopically from excitations of quantum fields, thence it should be possible to derive those from first principles within QFT on curved spacetimes. Such an analysis of the standard cosmological model within QFT on curved spacetimes has been performed by one of us in \cite{Hack:2013aya} and we shall review it in what follows.

A comprehensive analysis from this perspective could proceed as follows. One considers the full standard model of particle physics plus potential other fields and interactions as a perturbative interacting QFT on curved spacetime. One then aims to find a pair $(\omega,g)$, where $\omega$ is a Hadamard state on the algebra of this field model and $g$ is a metric on the manifold $M=I \times {\mathbb R}^3\subset {{\mathbb R}^4}$ of FLRW type, such that a) $(\omega,g)$ is a solution of the semiclassical Einstein equation
$$G_{ab}=8\pi G \omega(T_{ab})$$
where $T_{ab}$ is the energy momentum tensor of the field model and b) \eqref{eq_friedmannlcdm} are \eqref{eq_omegas} are satisfied up to suitably small corrections. Unfortunately such an analysis is quite involved, but we can consider a number of simplifications. First, we disregard all field interactions. This is a legitimate approximation if we consider the cosmological evolution only after the primordial synthesis of light nuclei -- the so-called Big Bang Nucleosynthesis (BBN) -- as field interactions are usually assumed to be irrelevant for the large-scale properties of the quantum state after this era. In the standard cosmological model, this enters by assuming that the each component of the perfect fluid in  \eqref{eq_friedmannlcdm} satisfies an individual conservation equation. As a further simplification, we disregard the spin of the quantum fields and model all massive fields, i.e. "matter", by a single massive scalar field, and all massless fields, i.e. "radiation", by a single massless scalar field, where both fields are considered to be conformally coupled to the scalar curvature ($\xi=\frac16$). This is done for ease of presentation as computations with higher spin fields are in principle straightforward, see for instance \cite{Dappiaggi:2009xj, Dappiaggi:2010gt}; the conformal coupling $\xi=\frac16$ is chosen because it simplifies computations and because the massless Dirac equation and the Maxwell equation are invariant under conformal isometries. Finally, provided one is able to assign a state $\omega$ to a FLRW metric $g$ in a coherent way, $\omega$ is in general a non-trivial functional of $g$ and thus obtaining an explicit solution of the semiclassical Einstein equation is at best difficult. In a recent yet unpublished work, Pinamonti and Siemssen have proven by a fixed point argument that the semiclassical Einstein equation can be uniquely solved for a linear scalar field model and a large class of initial conditions on a Cauchy surface, but for a quantitative analysis one needs to know the solution explicitly. We thus solve the semiclassical Einstein equation in the following approximate sense. We assume that the FLRW spacetime is given and determined by \eqref{eq_friedmannlcdm} and \eqref{eq_omegas}. On this spacetime we seek to find a pair of quantum states $\omega^m$ and $\omega^0$ for the massive and massless scalar field such that the sum of the energy densities in this states satisfies 
\begin{equation}\label{eq_approxsol}\frac{\omega^0(\rho) + \omega^m(\rho)}{\rho_0}=\Omega_\Lambda + \frac{\Omega_m}{a^3} + \frac{\Omega_r}{a^4}=\frac{\rho_\text{$\Lambda$CDM}}{\rho_0}\end{equation}
and \eqref{eq_omegas} up to suitably small corrections in the time interval of interest $z\in \left[0,10^9\right]$, where $z=0$ marks the present and $z=10^9$ is the redshift at which BBN took place.

In order to follow this program, it is useful to have at ones disposal a map which assigns a state $\omega$ to a FLRW metric $g$ {\it in a given coordinate system}; indeed this is necessary in order for the semiclassical Friedmann equation $3H^2=8\pi G\omega(\rho)$ to be well-defined in the first place. Such a construction is provided by the so-called states of low energy introduced by Olbermann \cite{Olbermann:2007gn}. These states minimize the energy density integrated in (cosmological) time with a sampling test function $f$ and are pure, Gaussian, isotropic and homogeneous states of Hadamard type. Their two-point Wightman function is (barring an $\epsilon$-prescription) of the form
$$\omega(x,y)=\frac{1}{8\pi^3 a(\tau_x)a(\tau_y)}\int\limits_{\mathbb{R}^3} d{\vec k}\, \overline{\chi_k(\tau_x)}\chi_k(\tau_y)e^{i\vec{k}(\vec{x}-\vec{y})}\,,$$
where the modes $\chi_k$ satisfy the ordinary differential equation
\begin{equation}\label{eq_modesode}
\left(\partial^2_\tau+k^2+m^2a^2 + \left(\xi-\frac16\right)R a^2\right)\chi_k(\tau)=0
\end{equation}
and the normalisation condition
\begin{equation}\label{eq_modesnormal}
{\chi_k}\partial_\tau \overline{\chi_k}-\overline{\chi_k}\partial_\tau{\chi_k}=i\,.
\end{equation}
Here, $k= |\vec{k}|$ and $\overline{\cdot}$ denotes complex conjugation. The modes $\chi_k$, which determine the state, are obtained by choosing arbitrary but fixed reference modes. The Bogoliubov coefficients in this mode basis are suitable functionals of the reference modes and the sampling function $f$. Olbermann has proven the Hadamard property of these states only for the case $\xi = 0$, but one can show that they are at least sufficiently regular in order to compute the energy density also in the case $\xi=\frac16$. If $\xi=\frac16$ and $m=0$, then the Hadamard property follows from the fact that these states are related to the Minkowski vacuum state by a conformal isometry. In the following, we set $\xi=\frac16$. A further assignment of a state to a FLRW spacetime in a given coordinate system is given by the so-called adiabatic states of order 0 introduced in \cite{Parker} and further developed in \cite{Luders, Junker}. These are defined by the modes which satisfy \eqref{eq_modesode} and the initial conditions $\chi_k(\tau)|_{\tau=\tau_0}=\widetilde{\chi}_k(\tau)|_{\tau=\tau_0}$, $\partial_\tau\chi_k(\tau)|_{\tau=\tau_0}=\partial_\tau \widetilde{\chi}_k(\tau)|_{\tau=\tau_0}$, where 
\begin{equation}\label{eq_adiabatic}\widetilde{\chi}_k(\tau)=\frac{1}{\sqrt{W(k,\tau)}}\exp\left({-i \int^\tau_{\tau_0}W(k,\tau^\prime) d\tau^\prime}\right),\quad W(k,\tau)=\sqrt{k^2+m^2a^2}.\end{equation}
The functions $\widetilde{\chi}_k(\tau)$ solve \eqref{eq_modesnormal} exactly but \eqref{eq_modesode} only approximately with error terms quantified by $\frac{Hm }{W^2}$ and $\frac{\partial_\tau Hm}{W^3}$. A detailed discussion of the error terms can be found in \cite{Olver}.

In the $\Lambda$CDM model, the radiation contribution $\frac{\Omega_r}{a^4}$ to the energy density is mostly of thermal nature, while the matter contribution $\frac{\Omega_m}{a^3}$ is mostly due to Dark Matter, which in some scenarios is believed to be of thermal origin as well. Motivated by this we look for states which satisfy \eqref{eq_approxsol} and \eqref{eq_omegas} among suitable ``thermal excitations'' of states of low energy. A fully satisfactory generalisation of the concept of thermal equilibrium to general curved spacetimes or even FLRW ones does not exist so far. Probably the most elaborated idea is the so-called local thermal equilibrium approach, see e.g. \cite{VerchRegensburg, Solveen:2012ai} for a review. Here we take a more pragmatic approach and consider the states introduced in \cite{Dappiaggi:2010gt}. Given a pure, Gaussian, isotropic and homogeneous Hadamard state $\omega$ specified by modes $\chi_k$, one can construct a family of Gaussian Hadamard states $\omega_{\beta, a_F}$ by defining the two-point Wightman function (up to an $\epsilon$-prescription) as
\begin{equation}\label{eq_genth}\omega(x,y)=\frac{1}{8\pi^3 a(\tau_x)a(\tau_y)}\int\limits_{\mathbb{R}^3} d{\vec k}\;e^{i\vec{k}(\vec{x}-\vec{y})} \left(\frac{\overline{\chi_k(\tau_x)}\chi_k(\tau_y)}{1-e^{-\beta k_0}}+\frac{\chi_k(\tau_x)\overline{\chi_k(\tau_y)}}{e^{\beta k_0}-1}\right)\,,\end{equation}
with $k_0:=\sqrt{k^2+m^2a^2_F}$. If $\chi_k$ are the modes of a state of low energy, these states match the almost equilibrium states introduced by K\"usk\"u in \cite{Kusku:2008zz} up to the form of $k_0$. The Hadamard property of the states defined by \eqref{eq_genth} follows from results of \cite{Pinamonti:2010is}. In the massless case, these states are independent of $a_F$ and satisfy the conformal KMS condition with respect to the conformal Killing vector $\partial_\tau$. In the massive case, they are considered to describe approximately the quantum state of a field which has been in thermal equilibrium in the distant past, and has ``frozen out'' of equilibrium at the time $a=a_F$. This corresponds to the phenomenological picture behind Dark Matter of thermal origin in the standard literature see e.g. \cite{Kolb:1990vq}.

Given this choice of quantum states we are left with the cumbersome task to compute the energy density in these states. To this avail, we can rewrite the singular part $h^\text{sing}_N(x,y)$ of a Hadamard solution    in terms of a Fourier integral in order to match the mode expansion of the states at hand, see \cite{Eltzner:2010nx, Pinamonti:2010is, Schlemmer, Degner}. In this way we obtain a Fourier integral expression for the regular part $w^h_N(x,y)$ of the relevant two-point Wightman function. The energy density is obtained by applying to this regular object a second order bi-differential operator and then taking the limit $x\to y$, cf. \eqref{eq_deftmunu}. This is well-defined and independent of $N$ if $N\ge 1$. As a result, we obtain the energy density as a convergent integral over $k$. In the massless case, this integral can be computed analytically. In the massive case however, both the integrand and the integral have been computed in \cite{Hack:2013aya} partly numerically and partly using analytical approximations. The reasons are manifold. To name a few, the mode equation \eqref{eq_modesode} can not be solved analytically on FLRW spacetimes of the form \eqref{eq_friedmannlcdm} if $m>0$. Moreover, even a numerical solution fails to be feasible for $m\gg H_0$ -- which is the realistic case as $H_0\simeq 10^{-33}$eV -- because the modes oscillate heavily. To overcome the latter problem the approximate adiabatic modes $\widetilde{\chi}_k(\tau)$, cf. \eqref{eq_adiabatic}, have been used as reference modes for the computation of the modes of the state of low energy, as they approximate the exact adiabatic modes of order zero particularly well exactly in the the regime $m\gg H$. 

Altogether the following results can be obtained. To discuss these, we rewrite the total energy density of the massless and massive conformally coupled scalar fields in the respective generalised thermal states \eqref{eq_genth} defined with respect to states of low energy as follows
\begin{equation}\label{eq_totaled}\frac{\omega^0(\rho) + \omega^m(\rho)}{\rho_0}=\frac{\rho^m_\text{gvac}+\rho^0_\text{gvac}+\rho^m_\text{gth}+\rho^0_\text{gth}}{\rho_0}+\gamma \frac{H^4}{H^4_0}+\Omega_\Lambda+\delta\frac{H^2}{H^2_0}+\epsilon\frac{J_{00}}{H^4_0}\;.\end{equation}

$\Omega_\Lambda$, $\delta$ and $\epsilon$ parametrise the freedom in the definition of the energy density as per \eqref{eq_nonunique}. The number of free parameters in this equation has been reduced to three, because $I_{ab}$ and $J_{ab}$ are proportional in FLRW spacetimes. We take the point of view that $\delta$, which effectively renormalises Newton's constant, is not a free parameter because Newton's constant has been measured already. In order to do this, we have to fix a value for the inverse length scale $\mu$ in the singular part of a Hadamard solution $h^\text{sing}_N(x,y)$, we do this by confining $1/\mu$ to be a scale in the range in which the strength of gravity has been measured. Because of the smallness of the Planck length, the actual value of $1/\mu$ in this range does not matter as changing $1/\mu$ in this interval gives a negligible contribution to the energy density. One could also take a more conservative point of view and consider $\delta$ to be a free parameter, in this case comparison with cosmological data, e.g. from Big Bang Nucleosynthesis, would presumably constrain $\delta$ to be very small once $1/\mu$ is in the discussed range.

On this occasion, we would like to highlight the point of view on the so-called cosmological constant problem taken here, as well as in most works on QFT on curved spacetimes in the algebraic approach and e.g. the review \cite{Bianchi:2010uw}. It is often said that QFT {\it predicts} a value for the cosmological constant $\Lambda$ and thus for $\Omega_\Lambda$ which is way too large in comparison to the one measured. This conclusion is reached by computing one or several contributions to the vacuum energy in Minkowski spacetime $\Lambda_\text{vac}$ and finding them all to be too large, such that, at best, a fine-tuned subtraction in terms of a negative bare cosmological constant $\Lambda_\text{bare}$ is necessary in order to obtain the small value $\Lambda_\text{vac}+\Lambda_\text{bare}$ we observe. Here, we assume the point of view that it is not possible to provide an {\it absolute} definition of energy density within QFT on curved spacetimes, and thus neither $\Lambda_\text{vac}$ nor $\Lambda_\text{bare}$ have any physical meaning by themselves; only $\Lambda_\text{vac}+\Lambda_\text{bare}$ is physical and measurable and any cancellation which happens in this sum is purely mathematical. The fact that the magnitude of $\Lambda_\text{vac}$ depends on the way it is computed, e.g. the loop or perturbation order, cf. e.g. \cite{Sola:2013gha}, is considered to be unnatural following the usual intuition from QFT on flat spacetime. However, it seems more convincing to us to accept that $\Lambda_\text{vac}$ and $\Lambda_\text{bare}$ have no relevance on their own, which does not lead to any contradiction between theory and observations, rather than the opposite. In the recent work \cite{Holland:2013xya} it is argued that a partial and unambiguous relevance can be attributed to $\Lambda_\text{vac}$ by demanding $\Lambda_\text{bare}$ to be analytic in all coupling constants and masses of the theory; taking this point of view, one could give the contribution to $\Lambda_\text{vac}$ which is non-analytic in these constants an unambiguous meaning. Indeed the authors of \cite{Holland:2013xya} compute a non-perturbative and hence non-analytic contribution to $\Lambda_\text{vac}$, which turns out to be small. In the view of this, one could reformulate the above statement and say that contributions to $\Lambda_\text{vac}$ and $\Lambda_\text{bare}$ which are analytic in masses and coupling constants have no physical relevance on their own.

The term in \eqref{eq_totaled} proportional to $\gamma$, which is not present in the $\Lambda$CDM-model, appears due to the so-called  trace anomaly, which is a genuine quantum and state-independent contribution to the quantum energy momentum tensor, see e.g. \cite{Wald3}. This term is fixed by the field content, $\gamma\simeq 10^{-122}$ for two scalar fields. As $H< H_0 z^2$ in the $\Lambda$CDM-model for large redshifts, this term can be safely neglected for $z<10^9$. 

The first terms in \eqref{eq_totaled} denote the genuinely quantum state dependent contributions to the energy densities of the two quantum fields. We have split these contributions into parts which are already present for infinite inverse temperature parameter $\beta$ in the generalised thermal states, and thus could be considered as contributions due to the states of low energy as generalised vacuum states ($\rho^m_\text{gvac}$, $\rho^0_\text{gvac}$), and into the remaining terms, which could be interpreted as purely thermal contributions ($\rho^m_\text{gth}$, $\rho^0_\text{gth}$). One can show that, up to the freedom parametrised by $\Omega_\Lambda$, $\delta$ and $\epsilon$, $\rho^0_\text{gvac}=0$ for arbitrary sampling functions $f$, whereas $\rho^m_\text{gth}/\rho_\text{$\Lambda$CDM}\ll 1$ for small masses $m\simeq H_0$ and large masses $m\gg H_0$ if the sampling function $f$ defining the state of low energy has sufficiently large support in time. This generalises results obtained by Degner on de Sitter spacetime \cite{Degner} and indicates that states of low energy with broad sampling functions are reasonable generalised vacuum states on FLRW spacetimes.

\begin{figure}
\includegraphics[width=1\columnwidth]{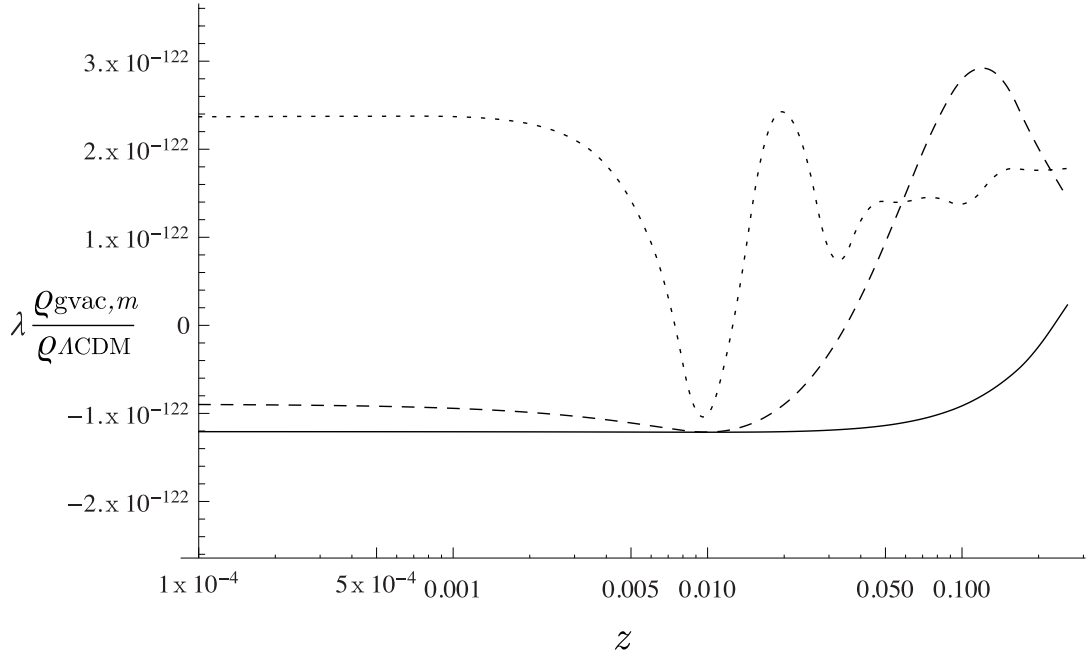}
\caption{\label{fig_rhonormsmallz}$\lambda \rho^m_\text{gvac}/\rho_\text{$\Lambda$CDM}$ for $z<1$ for various values of $m$ (rescaled for ease of presentation). The dotted line corresponds to $m=100H_0$ and $\lambda=10^{-2}$, the dashed line to $m=10H_0$ and $\lambda=1$ and the solid line to $m=H_0$ and $\lambda=10^{2}$. One sees nicely how the energy density is minimal in the support of the sampling function at around $z=10^{-2}$.}
\end{figure}

As for the thermal contributions, one finds in the massless case $$\rho^0_{\text{gth}}=\frac{\Omega_r}{ a^4}\quad \text{with}\quad \Omega_r=\frac{\pi^2}{30 \beta^4}.$$ Up to degree of freedom factors, this gives the $\Lambda$CDM value $\Omega_r \simeq 10^{-4}$ if the temperature parameter $1/\beta$ is in the range of the   Cosmic Microwave Background temperature $1/\beta \simeq 2.7$K. In the massive case, one can take typical values of $\beta$, $a_F$ and $m$ from Chapter 5.2 in \cite{Kolb:1990vq} computed by means of effective Boltzmann equations. A popular candidate for Dark Matter is a weakly interacting massive particle (WIMP), e.g. a heavy neutrino, for which \cite{Kolb:1990vq} computes
$$x_F = \beta a_F m \simeq 15 + 3\log(m/\text{GeV}) \qquad a_F \simeq 10^{-12}(m/\text{GeV})^{-1}\,.$$
Using this one finds for large $m$
$$\rho^m_\text{gth}\simeq \frac{1}{(2\pi)^{3/2}}\frac{m}{\beta^3 a^3}x^{\frac32}_F e^{-x_F}\,,$$
and thus $\Omega_m\simeq 0.3$ for $m\simeq 100$GeV.

At this stage, we have already seen that there exist states for the field model under consideration for which the energy density in the time interval $z\in[0,10^9]$ is of the form
\begin{equation}\label{eq_final}\frac{\omega^0(\rho) + \omega^m(\rho)}{\rho_0}=\Omega_\Lambda + \frac{\Omega_m}{a^3} + \frac{\Omega_r}{a^4} + \epsilon \frac{J_{00}}{H^4_0}\end{equation}
with $\Lambda$CDM values for $\Omega_m$, $\Omega_r$ and $\Omega_\Lambda$. This is the desired result up to the term $\epsilon \frac{J_{00}}{H^4_0}$ which is not present in the $\Lambda$CDM model, but quantified by the free parameter $\epsilon$. To analyse the influence of this term, we solve the equation
\begin{equation}\label{eq_extended}\frac{H^2}{H^2_0}=\Omega_\Lambda + \frac{\Omega_m}{a^3} + \frac{\Omega_r}{a^4} + \epsilon \frac{J_{00}}{H^4_0}\,.\end{equation} As $J_{00}$ contains second derivatives of $H$, this equation can be rewritten as a second order ordinary differential equation for $H(z)$ and solved by choosing e.g. $\Lambda$CDM initial conditions at $z=0$. This analysis is consistent as the derivation of \eqref{eq_final} does not only hold for $\Lambda$CDM-backgrounds \eqref{eq_friedmannlcdm}, but also for backgrounds of the form \eqref{eq_extended}. One finds that for large redshifts $z$, the solution of \eqref{eq_extended} is of the form
$$\frac{H^2}{H^2_0}=\Omega_\Lambda + \frac{\Omega_m}{a^3} + \frac{\widetilde{\Omega_r}(\epsilon)}{a^4}$$
with $\widetilde{\Omega_r}(\epsilon)\ge \Omega_r$, thus the term $\epsilon J_{00}$ effectively generates additional energy density of radiation type in the early universe, i.e. Dark Radiation. Surprisingly, one finds $\lim_{\epsilon \downarrow 0}\widetilde{\Omega_r}(\epsilon)=\Omega_r$, but $\lim_{\epsilon \uparrow 0}\widetilde{\Omega_r}(\epsilon)=\infty$. This is well in line with earlier results on the stability of the Einstein equation with additional higher order derivative terms, e.g. \cite{Anderson1, FW96, Haensel, Koksma, ParkerSimon, Star}. The value of $\widetilde{\Omega_r}$ can be constrained by observations of the primordial fractions of light nuclei as predicted by BBN, since the synthesis of these nuclei depends sensitvely on the Hubble rate at $z\simeq 10^9$. It turns out that $\widetilde{\Omega_r}(\epsilon)$ is in conflict with observations for $\epsilon < 0$, but that the BBN data are compatible with $0\le \epsilon < 2\times 10^{-15}$ if all Dark Radiation is attributed to the origin discussed here.

The value of $\epsilon$ can be constrained also by other means. On the one hand, a further bound on $\epsilon$ can be obtained by analysing the effects of higher derivative contributions to the gravitational Lagrangean in the context of Inflation. In fact, an early inflationary model proposed by Starobinsky in \cite{Starobinsky:1983zz} is based on an $\epsilon J_{00}$ contribution to the energy density. Confronting this inflationary model with current Cosmic Microwave Background data yields $\epsilon\simeq 10^{-113}$ \cite{Kaneda:2010ut}. Thus, if Inflation occurred due to the $\epsilon J_{00}$ contribution to the energy density, then $\epsilon$ is too small for generating a considerable amount of Dark Radiation. However, if Inflation has a different origin or did not occur at all, then one obtains the lower bound $\epsilon > 10^{-113}$. Finally, an upper bound on $\epsilon$ can be obtained by considering the Newtonian limit of the semiclassical Einstein equation. In this limit, the higher order derivative terms $I_{ab}$ and $J_{ab}$ in \eqref{eq_nonunique} generate two Yukawa corrections to the Newtonian potential of a point mass of opposite sign \cite{Stelle:1977ry}. Assuming that these corrections don't cancel on the relevant length scales, one can obtain bounds on the strength and typical length scale of these Yukawa terms from torsion-balance experiments \cite{Kapner:2006si} and consequently the upper bound $\epsilon < 10^{-60}$ \cite{Calmet:2008tn}. Again, this upper bound would imply that $\epsilon$ is too small for generating a considerable amount of Dark Radiation. However, it is still possible that the aforementioned Yukawa corrections cancel each other on the length scales relevant for the experiments described in \cite{Kapner:2006si}, such that $\epsilon$ could be as large as our upper bound, which in this case would give a real bound on one and hence both Yukawa corrections. Moreover, the bounds inferred from \cite{Kapner:2006si} and from the analysis reviewed here stem from phenomena on completely different length scales. As a rough estimate we note that the diameter of our observable universe, which today is about $6/H_0\simeq 10^{27} $m, was at e.g. $z=10^9$ still $10^{18}$m and thus much larger than the submillimeter scales relevant for the torsion-balance experiments. Thus it could be that effects we have not considered yet, e.g. state-dependent effects which are due to the small-scale structure of the quantum states we have fixed only on cosmological scales so far, affect the comparison between the two different sources of input for the determination of $\epsilon$. 

We conclude that a more fundamental understanding of the standard cosmological model appears to be possible within QFT on curved spacetimes. In this framework one even finds a new free parameter not present in the standard model. This parameter can potentially account for Dark Radiation, the existence and nature of which are currently topics of active research.

\end{document}